# Optical gears in a nanophotonic directional coupler

Fengchun Zhang[†,§,‖,⊥,#], Yao Liang[‡,#], Heran Zhang[†,§,⊥], Yong Zhang[§,‖], Xu-Guang Huang[*,†,§,⊥], Baohua Jia[*,‡] and Songhao Liu[†,§,‖,⊥]

[†]Guangzhou Key Laboratory for Special Fiber Photonic Devices and Applications, South China Normal University, Guangzhou, 510006, China.

[‡]Centre for Micro-Photonics, Faculty of Science, Engineering and Technology, Swinburne University of Technology, Hawthorn, Victoria 3122, Australia.

[§] Guangdong Provincial Key Laboratory of Nanophotonic Functional Materials and Devices, South China Normal University, Guangzhou, 510006, China.

[‖]Guangdong Engineering Research Center of Optoelectronic Functional Materials and Devices, Institute of Opto-Electronic Materials and Technology, South China Normal University, Guangzhou, 510631, China.

[⊥]Specially Functional Fiber Engineering Technology Research Center of Guangdong Higher Education Institutes, Guangdong Provincial Engineering Technology Research Center for Microstructured Functional Fibers and Devices, South China Normal University, Guangzhou, 510006, China

**ABSTRACT:** Gears are rotating machines, meshing with each other by teeth to transmit torque. Interestingly, the rotating directions of two meshing gears are opposite, clockwise and counterclockwise. Although this opposite handedness motion has been widely investigated in machinery science, the analogue behavior of photons remains undiscovered. Here, we present a simple nanophotonic directional coupler structure which can 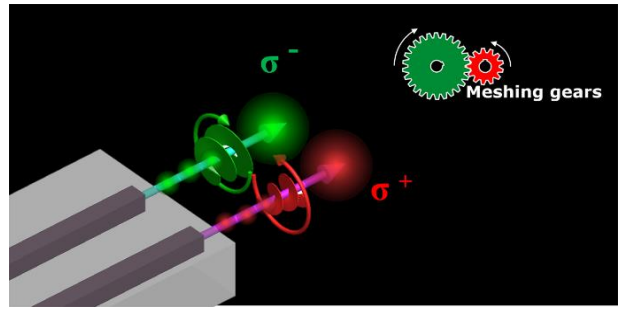 generate two meshing gears of angular momentum (AM) of light—optical gears. Due to the abrupt phase shift effect and birefringence effect, the AM states of photons vary with the propagation distance in two adjacent waveguides of the coupler. Thus, by the choice of coupling length, it is able to obtain two light beams with opposite handedness of AM, confirming the appearance of optical gears. The full control in the handedness of output beams is achieved via tuning the relative phase between two orthogonal modes at the input ports. Optical gears thus offer the possibility of exploring light-matter interactions in nanoscale, opening up new avenues in fields of integrated quantum computing and nanoscale bio-sensing of chiral molecules.

**KEYWORDS:** *optical gears, angular momentum, abrupt phase shift, nanophotonic waveguides, spin and orbit interactions*

The abrupt phase shift is a fundamental phenomenon in many classical and quantum resonant systems where energy exchange is possible, such as RLC circuits,[1] coupled pendulums[2] and quantum dots (QDs),[3] as shown in Figure 1a. In optics, this phenomenon has been observed both in bulk and nanophotonic systems, i.e., interface reflections (Figure 1b),[4] metasurfaces[5-7] and directional couplers.[8] The abrupt shift has been recently gathering increasing interest, as it plays an important role in many light-matter interactions with exotic effects, such as negative refraction and reflection,[9] photonic spin Hall effect,[10] spin-orbit coupling[11] and chiral beam distinguishing.[12, 13]

In particular, this abrupt phase shift occurs when light coupling in a directional coupler consisting of two photonic waveguides (WGs). Already a number of integrated devices have been realized based on directional couplers, such as optical filters,[14] 3-dB splitters,[15] polarization beam splitters,[16, 17] PT-symmetric nonlinear couplers,[18] entangled photon-pairs sources,[19] two-photon quantum interference,[20] integrated quantum logical gates,[21] and all-optical data processing.[22]

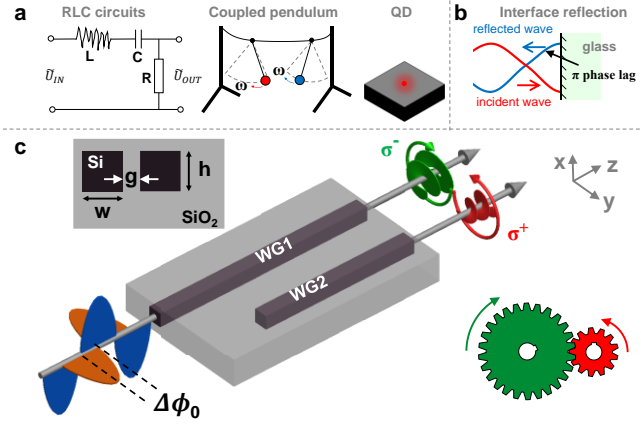

Figure 1 (a) The abrupt phase shift found in classical and quantum systems. (b) A π phase lag observed in the reflection of an interface between glass and air. (c) Schematic of the proposed structure with geometric details. The optical analogue of two meshing gears: the output beams have the opposite handedness of AM states. The coordinate system used.

However, most of those demonstrations have been focused on a single quasi-linearly polarized mode and mainly discussed the energy exchange between two adjacent waveguides. Innovations concerning multi-polarized modes, i.e. quasi-circularly (or elliptically) polarized modes,[23] and the abrupt π phase shift, which is introduced by the coupling process, remain largely unexplored.

In this work, we show that they have much potential for creations of novel devices as well. For example, a directional coupler is a crucial ingredient for the manipulation of angular momentum (AM) of light in nanophotonic waveguides when two orthogonal polarized modes are involved. By engineering the length of coupling region, it is possible to construct an optical analogue of two meshing gears, where the quasi-elliptically polarized modes have opposite handedness in two adjacent waveguides at the output ports. To our knowledge, it is the first time that this new concept of optical gears is proposed. In addition, the handedness of the output modes can be manipulated via the choice of the relative phase ($\Delta\phi_0$) of quasi-TE and -TM modes at the input port (Figure 1c). Interestingly, our scheme is conceptually different from previous methods for manipulation of AM, such as birefringence effect caused by optical crystals[24] and abrupt phase change introduced by nano-resonators.[25] Instead, we show that, in the coupling region of a directional coupler, the phase lag between the two orthogonally polarized modes is modulated via two factor: the abrupt phase shift and the birefringence effect that happens in the coupling process.

## RESULTS AND DISCUSSION

The proposed scheme is sketched in Figure 1c. The directional coupler consists of two uniform parallel silicon (Si) waveguides. The width (w) and height (h) of each waveguide are identical, w = h = 340 nm, and the gap between them is g = 40 nm. We assume the whole structure is surrounded by silica ($SiO_2$) and the operating wavelength is 1.55 μm.

We first discuss the abrupt phase shift and the birefringence effect in the coupler, and then the opposite handedness of AM behavior. As light propagates along the coupler, it couples from the first waveguide (WG1) to the second one (WG2) and then couples back to the first one again. By using the coupled mode approach, the light field dynamics of the coupling region is described by

$$\begin{cases} \dfrac{da_1(z)}{dz} = -in_1 k_0 a_1(z) + \kappa a_2(z) \\ \dfrac{da_2(z)}{dz} = -in_2 k_0 a_2(z) + \kappa a_1(z) \end{cases} \quad (1)$$

where $a_{1,2}(z)$ represent respectively the complex amplitudes of the light in the WG1 and WG2, while $\kappa = \pi/(2z_c)$ is the coupling coefficient with the coupling length $z_c$, $k_0 = 2\pi/\lambda$ the free space wavenumber, and $n_1 = n_2$ the effective indices. For our single-mode directional coupler, the light energy can be 100% exchanged between two waveguides, and eq 1 can be solved analytically,

$$\begin{pmatrix} a_1(z) \\ a_2(z) \end{pmatrix} = \begin{pmatrix} \cos(\kappa z) & -j\sin(\kappa z) \\ -j\sin(\kappa z) & \cos(\kappa z) \end{pmatrix} \begin{pmatrix} a_1(0) e^{-in_1 k_0 z} \\ a_2(0) e^{-in_2 k_0 z} \end{pmatrix} \quad (2)$$

assuming unit power entering the WG1 with the electric field of $\vec{E}_1 = e^{-i(\omega t - \beta z)} \vec{e}$, where β is the propagation constant. Correspondingly, the initial conditions are $a_1(0) = 1$ and $a_2(0) = 0$. For $0 < z < z_c$, eq 2 can be simply written as,

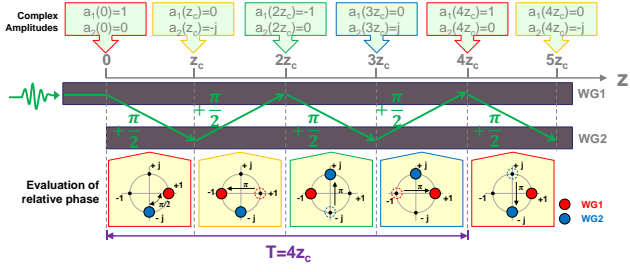

Figure 2. Schematic of the abrupt phase shift introduced in the coupling process. The complex amplitudes at the beginning position for each period (top). The relative phase evolution for each period, which is shown on a complex plane, and which includes the π phase shift information (bottom).

$$\begin{cases} a_1(z) = \cos(\kappa z) e^{-in_1 k_0 z} \\ a_2(z) = -j\sin(\kappa z) e^{-in_2 k_0 z} \end{cases} \quad (3)$$

The $-j$ term in eq 3 implies an intrinsic phase lag of $\pi/2$ for the light field in the WG2 compared with the one in the WG1. This solution is well described for the energy coupling process, but not sufficient for the description of the phase evolution. To make up this defect, it has to be modified by some mathematical transformations every time when light is totally coupled from one waveguide to another. Thus applying the mathematical transformations to eq 3 (see Supporting Information for details), a general description for the energy and phase evolution in the coupler can be written as

$$\begin{cases} a_1(z) = \cos(\kappa z - nz_c) e^{-i(n_1 k_0 z + n\frac{\pi}{2})} \\ a_2(z) = \sin(\kappa z - nz_c) e^{-i[n_2 k_0 z + (n+1)\frac{\pi}{2}]} \end{cases} \quad n = 0, 2, 4\ldots$$

$$\begin{cases} a_1(z) = \sin(\kappa z - nz_c) e^{-i[n_1 k_0 z + (n+1)\frac{\pi}{2}]} \\ a_2(z) = \cos(\kappa z - nz_c) e^{-i(n_2 k_0 z + n\frac{\pi}{2})} \end{cases} \quad n = 1, 3, 5\ldots \quad (4)$$

for $nz_c < z < (n+1)z_c$, where n is an integer. eq 4 is a periodic solution with a period $T = 4z_c$. In addition, for $z \in (2nz_c, (2n+1)z_c)$, the phase of light field in WG1 is $\pi/2$ in advance compared with the one in WG2, while an opposite situation occurs for $z \in ((2n-1)z_c, 2nz_c)$.

Besides, at every point where $z = (2n+1)z_c$, a π phase shift happens in WG1 and the abrupt phase shift (π) occurs in WG2 at the points where $z = 2nz_c$. To visualize this finding, at the bottom of Figure 2, we plot the evolution of relative phase between WG1 and WG2 in the complex plane for each period.

It should be emphasized that light coupling between waveguides is a resonance phenomenon, which is an analogy to the standing wave of a laser's resonant cavity consisting of two mirrors. There is also a π phase shift between the incident light and the reflected light when the light is reflected by the mirrors, which can be well explained by Fresnel equations.[26]

To discuss the birefringence effect in the coupler, we apply the supermode solution to analyse the coupling process. In the coupling region, the coupler can be regarded as a two cores waveguide, and the guided modes can be linearly represented by a symmetric (even, $\beta_+$) and an anti-symmetric (odd, $\beta_-$) modes. Usually, the propagation constants of the even and odd modes are not equal ($\beta_+ \neq \beta_-$) but with small difference. Thus their interference pattern results in a beat in the waveguides, with the beat length $z_b = 2z_c = 2\pi/(\beta_+ - \beta_-)$, where $\beta_{+,-} = n_{+,-}k_0$ are the propagation constants and $n_{+,-}$ are the effective indices of the even and odd modes. The average propagation constant for the light in each waveguide is $\bar{\beta} = (\beta_+ + \beta_-)/2$.

Considering both the propagation effect and abrupt phase change effect, for WG1, the phase of light at different longitudinal positions is thus given by,

$$\phi_1(z) = \bar{\beta}z + \pi\left(floor(\frac{z}{2z_c} + \frac{1}{2})\right) + \phi_1(0) \quad (z>0) \quad (5)$$

where floor(x) is the floor function such that floor(x) is the largest integer not greater than x, and $\phi_1(0)$ is the initial phase at the position z=0. The first term suggests that the light propagate along the +z direction while the second term indicates the abrupt phase shift (π) introduced by light coupling. Accordingly, the phase distribution of light for WG2 is,

$$\phi_2(z) = \frac{\pi}{2} + \bar{\beta}z + \pi\left(floor(\frac{z}{2z_c})\right) + \phi_1(0) \quad (z>0) \quad (6)$$

Nanophotonic silicon waveguides usually exhibit huge birefringence effect. However, in a rectangle Si waveguide surrounded by silica, where the width and height are equal, the propagation constants of the fundamental (zero order) quasi-TE and -TM modes are equal due to the diagonal symmetry, that is $\beta_{TE} = \beta_{TM}$. However, in the coupler, the even and odd modes for the

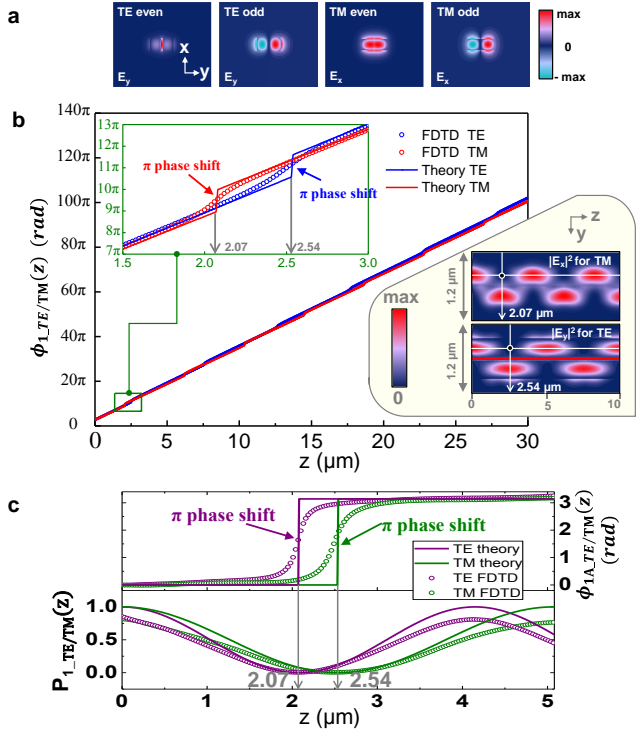

Figure 3 (a) mode distributions of the even and odd modes for the TE- and -TM polarized modes in the coupling region. (b) Theoretical (lines) and stimulated (symbols) dependences of the phase on the propagation distance (z). Inserted figure shows mode distributions of the dominant components at the yz-plane (cross the WGs centers) for the quasi-TE and -TM polarized modes. (c) Theoretical (lines) and stimulated (symbols) dependences of the abrupt phase shift and power (|E|2) on z in WG1.

TE- and TM-like polarized light are dramatically different from each other. Figure 3a displays the real parts of the dominant polarization components for two orthogonal polarized modes ($Re(E_x)$ for TM and $Re(E_y)$ for TE). Using the Eigenmodes Solver, which is available in finite-different time-domainate (FDTD) Solutions package from Lumerical Inc., the effective indices for the even and odd modes of TE-like polarized light are calculated to be 2.5637 and 2.2581, respectively, while the ones for TM-like polarized light are respectively 2.5311 and 2.1573. Thus, we have $\bar{\beta}_{TE} = 2.4109 k_0$ and $\bar{\beta}_{TM} = 2.3442 k_0$.

We perform computer FDTD simulation to confirm our theoretical analysis. In our simulation, we use the phase and power of the dominant polarization component ($E_x$ for TM and $E_y$ for TE) at the waveguide center point to represent the phase and power in each waveguide. Figure 3b shows the dependence of the phase on the propagation distance (z) for WG1. According to eq 5, the slopes of lines indicate the average propagation constants, $\bar{\beta}_{TE/TM}$, which are in good agreement with the simulation results. As for the power ($P \propto |E|^2$) of light, the normalized powers of the first and second waveguides have a characteristic given by

$$\begin{cases} P_1(z) = cos^2(\kappa z) \\ P_2(z) = sin^2(\kappa z) \end{cases} \quad (7)$$

where $\kappa = (\beta_+ - \beta_-) / 2$ is the coupling coefficient. Interestingly, a $\pi$ phase shift happens to both TE- and TM-polarized modes in the vicinity where the powers reach their minimum (0), as predicted by our abrupt phase shift theory. As an aid to comprehension, according to eq 5, we defined the abrupt phase shift term ($\phi_{1A}$) in WG 1 as

$$\phi_{1A}(z) = \phi_1(z) - \bar{\beta} z - \phi_1(0) \quad (z>0) \quad (8)$$

We plot the abrupt phase shift term and power dependence on the propagation distance (z) in Figure 3c. Although there are some minor disagreements between the analytical and stimulated results regarding the abrupt phase shift, the abruptness of $\pi$ phase shift is for sure for both of polarized modes. Also, this abruptness of $\pi$ phase shift is independent of the coupling length $z_c$ (Figure S1, Supporting Information). Thus, eq 5 and 6 are very good approximated methods for the prediction of phase of light in the coupler.

To investigate the evolution of angular momentum of light in the coupler, we respectively discuss the power and phase of light. We first assume a quasi-TE and -TM modes simultaneously entering the input port of WG1. These two orthogonal polarized modes will independently undergo different coupling processes in the coupler. As for the relative phase between the quasi-TE and -TM modes in the first and second waveguides, it is given by,

$$\Delta\phi_{1/2}(z) = \phi_{1/2\_TM}(z) - \phi_{1/2\_TE}(z) \quad (9)$$

Note that $e^{-i\Delta\phi_{1/2}(z)}$ are periodic functions, which means, $e^{-i\Delta\phi_{1/2}(z)} = e^{-i(\Delta\phi_{1/2}(z) \pm 2m\pi)}$, where m is an integer. We simplify the relative phase by omitting the redundant $2m\pi$. Thus, eq 9 could be written as,

$$\Delta\tilde{\phi}_{1/2}(z) = \mod\left[\phi_{1/2\_TM}(z) - \phi_{1/2\_TE}(z), 2\pi\right] \quad (10)$$

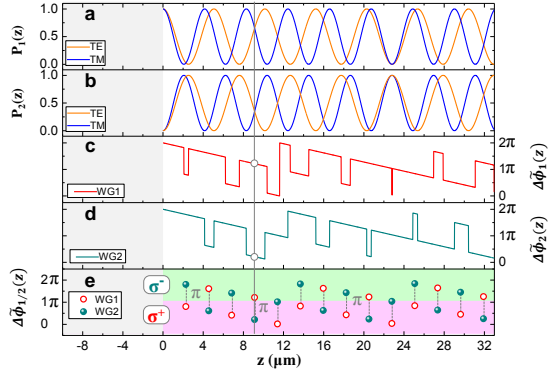
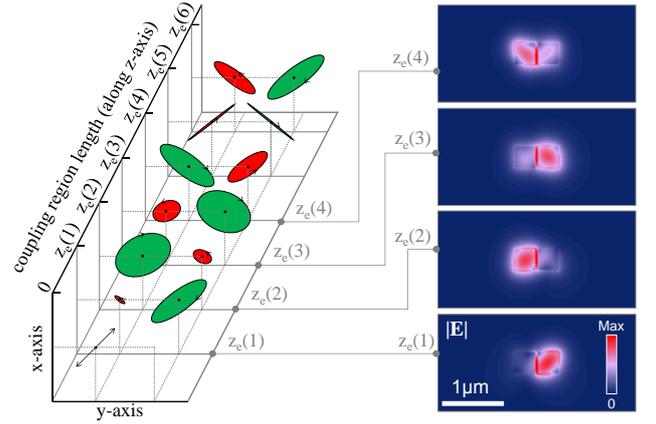

Figure 4. The theoretical results of powers dependence on the propagation distance (z) for the quasi-TE and -TM polarized modes in (a) WG1 and (b) WG2. The theoretical results of relative phase between the quasi-TE and -TM polarized modes, respectively in (c) WG1 and (d) WG2. (e) The theoretical results of relative phase between the quasi-TE and -TM polarized modes at the longitudinal positions $z_e$, where $P_{1,2\_TE}(z_e) = P_{1,2\_TM}(z_e)$ in WG1 and WG2.

Figure 5. Evolution of polarization at waveguides central points along the propagation length (z) of the coupler. The left side shows the theoretical prediction while the right side shows the simulated electric field distribution at the corresponding cross sections

where mod(A,B) is modulo operation that finds the remainder of A/B. Figure 4a-d show the theoretical powers and relative phases of the two polarized modes in the first and second waveguides, according to eq 7 and 10.

Although at first glance the relative phases $\Delta\tilde{\phi}_{1/2}(z)$ appear to be irregular, interestingly, we found that at some discrete positions where the amplitudes of the two polarized modes are equal ($P_{1\_TE}(z) = P_{1\_TM}(z)$, or $P_{2\_TE}(z) = P_{2\_TM}(z)$), the handedness of AM of photons in the first and second waveguides are perfectly opposite (Figure 4**e**). eq 7 indicates that the energy coupling ($P_1(z)$ and $P_2(z)$) is independent of the phase condition ($\phi_1(z)$ and $\phi_2(z)$), so that we can analyze the amplitudes and phases independently. At the point ($z_e$) where $P_{1\_TE}(z_e) = P_{1\_TM}(z_e)$, according to eq 7, it should satisfy

$$\cos^2(\kappa_{TM} z_e) = \cos^2(\kappa_{TE} z_e) \quad (11)$$

The solution to eq 11 is,

$$z_e(m) = \begin{cases} \dfrac{m\pi}{\kappa_{TM} + \kappa_{TE}} \\ \dfrac{m\pi}{\kappa_{TM} - \kappa_{TE}} \end{cases} \quad m=0,1,2,3 \quad (12)$$

where m is an integer. In our case, $z_e(m) \approx m*2.28$ μm or $z_e(m) \approx m*22.74$ μm. The latter case is relative large. In this work, we mainly discuss a coupler less than 10 μm, and thus, we neglect the effect caused by the latter solution. At these points, according to eq 5, 6, 9, 10 and 12 we have,

$$\left|\Delta\tilde{\phi}_1(z_e) - \Delta\tilde{\phi}_2(z_e)\right| = \pi \quad (13)$$

This eq 13 indicates that the handedness of polarization at two adjacent waveguides is precisely opposite.

To help comprehension of this optical meshing gears behavior (opposite handedness), in Figure 5, we plot the polarization states at the center points of two waveguides at these discrete positions. At these positions, usually a right-handed elliptically polarized mode at the left waveguide will accompany with a left-handed elliptically polarized mode at the right waveguide, and vice versa. This theoretical prediction is consistent with the FDTD simulation and the simulated electric field of each cross section is shown in the right side of Figure 5. Another interesting finding of the optical meshing gears behavior is that the center points polarizations of output modes can be steered by the choice of initial relative phase ($\Delta\phi_0$) between the quasi-TE and -TM modes at the input port. The proposed scheme for the control of $\Delta\phi_0$ is shown in the Supporting Information (**Figure S2**). Taking the fourth equal amplitude position (z = $z_e(4) \approx$ 9.12 μm) for example, a change in the initial phase leads to a change of the polarization of elliptical polarized modes in two waveguides at z = $z_e(4)$. Figure 6a shows the electric field distributions and center points polarizations of two waveguides at z = $z_e(4)$ for different $\Delta\phi_0$ value (0, $\pi/4$ and $\pi/2$). In addition, the relative phases ($\Delta\tilde{\phi}_{1/2}(z_e(4))$) show a linearly dependence on $\Delta\phi_0$ (Figure 6b). $\Delta\phi_0$ is independent of the energy coupling (Figure S3, Supporting Information). In particular, the opposite handedness characteristic of center points polarization at two adjacent waveguides still holds true for various $\Delta\phi_0$ values, since $|\Delta\phi_1 - \Delta\phi_2| = \pi$. This finding is important, as it allows for the manipulation of chirality of output modes. For example, it is able to obtain a right-handed (RH) quasi-circularly polarized mode in left waveguide while a left-handed (LH) one in the right for $\Delta\phi_0 \approx 0.29\pi$. Also, the choice of $\Delta\phi_0 \approx -0.71\pi$ leads to a LH quasi-circularly polarized mode in the left waveguide while a RH one in the right.

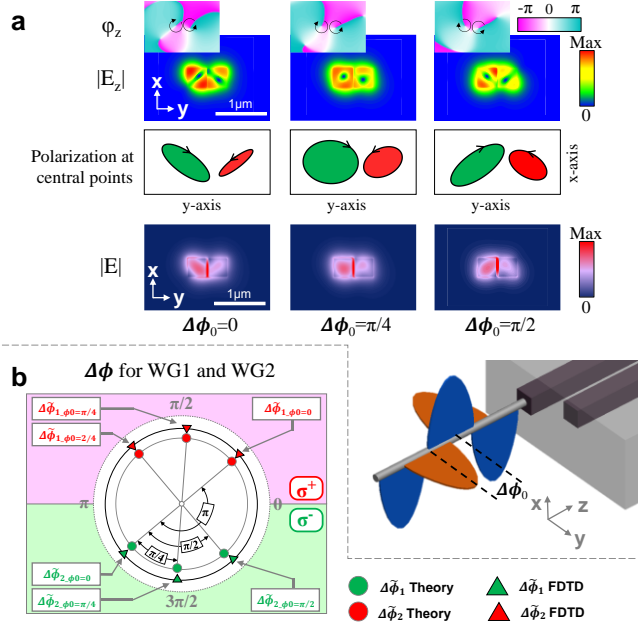

Figure 6. (a) The dependence of output modes at z = $z_e(4)$ on the initial relative phase $\Delta\phi_0$. The corresponding electric field distributions (|E| and |$E_z$|) and center points polarization states are shown. The spatial phase of $E_z$ indicates a phase singularity in each waveguide center for a quasi-elliptically polarized mode. (b) The dependence of relative phase ($\Delta\tilde{\phi}_{1/2}(z_e(4))$) of output modes on the initial relative phase $\Delta\phi_0$.

Surprisingly, the electric field distributions (|E|) of the three output modes are spatially different. In relative large waveguide structures, it is believed that the spatial mode distribution is independent of the polarization of light.[27] However, as the dimensions of waveguide scale down, these two quantities will inevitably get connected due to the spin-orbit interactions.[28] In other words, the polarization of light does indeed affect its spatial mode distribution in nanophotonic waveguides, which is our case.

Angular momentum can be decomposed into two components: the spin part (SAM) associated with the polarization and the orbital part (OAM) related to the spatial phase.[29] These two components can get coupled in nanophotonic waveguides. A quasi-elliptically polarized mode is usually accompanied by a longitudinal vortex component ($E_z$) due to the spin to orbital coupling in nanophotonic waveguides.[30,31] We find that the twisted handedness of the longitudinal component coincides with the center point polarization. For example, the center point of the left waveguide at z = $z_e(4)$ witnesses a right-handed elliptical polarization for $\Delta\phi_0 = \pi/4$, spinning clockwise. Accordingly, the longitudinal vortex component of the left waveguide at z = $z_e(4)$ twists clockwise, which is revealed by the spatial phase distribution of $E_z$ (Figure 6a).

We would like to emphasize the importance of this unique characteristic of opposite chirality, which is an optical analogue of two meshing gears transmitting rotational motion. Although some newly discovered optical phenomena, such as photonic wheels,[32] polarization of Möbius strips[33] and surface plasmon drumhead modes,[34] are often limited by immediately practical applications at the beginning, they may trigger increasing discussions later since they are strongly connected to fundamental physics and a variety of potential applications. It is therefore not unrealistic to expect that the optical gears phenomenon may open up new avenues in various fields, such as the integrated quantum science and on-chip chiral molecules detections, where the handedness of AM of photons is a key requirement.

## CONCLUSION

Our results uncover an exotic chirality phenomenon buried under the coupling process in a nanophotonic directional coupler, which has not been previously reported in literature. Namely, we introduce a new idea of optical gears where opposite handedness of AM can be obtained via a simple coupler structure, and where the chirality of AM is tunable via the choice of initial relative phase between two orthogonal modes at the input port. Also, we find that the polarization of modes vary along with the propagation distance in the coupler when two orthogonal mode involved, because of the abrupt phase shift effect and birefringence effect. The demonstration of a simple coupler capable of processing complex light beams carrying AM may open many possibilities for applications in fields ranging from fundamental physics and devices designing. For example, it could be applied to make an entangled photon source with tunability, for which the handedness of AM state is an available degree of freedom to encode quantum information.

## METHODS

In this work, the numerical experiments were carried out by utilizing a software named FDTD Solutions (A high performance 3D FDTD-method Maxwell solver from Lumerical Inc.). The boundary conditions in all the simulations process are chosen to be perfectly matching layer (PML). The simulation domain was divided into congruent cubes one another in all directions, and the mesh spacing in transverse (x and y directions) and longitudinal (z direction) sections are all 10 nm. The wavelength in all the theoretical calculations and simulation processes is set to be $\lambda$ = 1.55 μm. Correspondingly, the permittivities of silicon (Si) and silica ($SiO_2$) using in this work are 12.085 and 2.0851 respectively.[35] In addition, the effective mode indices of different structure sections are calculated using mode solver simulations (FDTD Solutions mode source). Also, the mesh spacing employed for transverse (x and y directions) sections is 10 nm.

## ASSOCIATED CONTENT

### Supporting Information

The Supporting Information is available free of charge on the ACS Publications website at DOI: 10.1021/acsphotonics.XXXXXXX.

Additional details (PDF).

## AUTHOR INFORMATION

**Corresponding Authors**
*E-mail: huangxg@scnu.edu.cn.
*E-mail: bjia@swin.edu.au.
**Author Contributions**
#F. Zhang and Y. Liang contributed equally to this work.
**Notes**
The authors declare no competing financial interest.

## ACKNOWLEDGMENTS

This work was supported by the Nature Science Foundation of China (61574064), the Project of Discipline and Specialty Constructions of Colleges and Universities in the Education Department of Guangdong Province (2013CXZDA012), Guangdong Natural Science Foundation (2014A030313446), the Program for Changjiang Scholars and Innovative Research Team in University (IRT13064), the Science and Technology Program of Guangdong Province (2015B090903078), and the Science and Technology Planning Project of Guangdong Province (2015B010132009).